# Influence of barrier form on Fowler-Nordheim plot analysis

Running title: Effect of barrier form on Fowler-Nordheim plot analysis

Running Authors: Fischer et al.

Andreas Fischer[a), b)], Marwan S. Mousa

Department of Physics, Mu'tah University, Al-Karak 61710, Jordan

Richard G. Forbes[c), d)]

Advanced Technology Institute & Dept. of Electronic Engineering, Faculty of Engineering and Physical Sciences, University of Surrey, Guildford, Surrey GU2 7XH, UK

[a)] Present address: Institut für Physik, Technische Universität Chemnitz, Chemnitz, Germany.

[b)] Electronic mail: andreas.fischer@physik.tu-chemnitz.de

[c)] Electronic mail: r.forbes@trinity.cantab.net

[d)] American Vacuum Society member.

**Abstract**

Recent research has described an improved method of Fowler-Nordheim plot analysis, based on the definition and evaluation of a slope correction factor and a new form of intercept correction factor. In this improved approach there exists a basic approximation that neglects certain terms in the general theory, and focuses on the influence of the form of the tunneling barrier on the values of basic slope ($\sigma_B$) and intercept ($\rho_B$) correction factors. Simple formulae exist that allow these to be evaluated numerically for a barrier of arbitrary well-behaved form. This paper makes an initial exploration of the effects of barrier form on FN plot analysis. For a planar emitter, two models for the correlation-and-exchange (C&E) potential energy (PE) are used. For the Schottky-Nordheim barrier, it is shown that numerical and analytical approaches generate equivalent results. This agreement supports the validity of the numerical methods used. Comparisons with results for the Cutler-Gibbons barrier show that small differences in the assumed C&E PE make little difference to values of $\sigma_B$ and $\rho_B$. Schottky's planar image PE has then been used, in conjunction with the electrostatic PE variation associated with a spherical emitter model, to explore the influence of apex radius $r_a$ on correction-factor values, for values of $r_a \geq 20$ nm. Both $\sigma_B$ and $\rho_B$ increase significantly as $r_a$ decreases, especially $\rho_B$. At low values of barrier field $F$, $\sigma_B$ depends approximately linearly on $1/F$, with a slope that depends on $r_a$. Suggestions are made for how the exploratory work described in this paper might be extended.

# I. INTRODUCTION

Fowler-Nordheim (FN) plots[1] are often used to interpret current-voltage characteristics related to cold field electron emission (CFE). This paper is one of several that will explore an improved approach to FN plot analysis. Specifically, it follows up a recent paper[2] that sets out the principles and high-level theory of an improved version of the tangent method[3,4] of analyzing FN plots. In this method, slope and intercept correction factors are used to obtain quantitatively more reliable values of parameters (such as field enhancement factor and emission area) that are extracted from the slope and intercept of FN plots. The present paper begins to explore (within the context of the "basic approximation" set out in Ref. 2) the effects of barrier form on the "basic" slope and intercept correction factors $\sigma_B$ and $\rho_B$ defined in Ref. 2.

Current thinking is that—when no "saturation" or other major perturbing effects are operating—corrections due to barrier form may be the most significant corrections to FN plot analysis. This thinking is necessarily provisional, because the high-level theory in Ref. 2 shows that in principle the improved theory contains several forms of correction, some of which have never been investigated in detail. It is also likely that saturation or other major perturbing effects often do operate, particularly for some classes of emitter. However, it is convenient to start by investigating barrier-form effects. One certainly expects the related theory to apply to metal emitters (or other good conductors) that are securely mounted, with a good conducting path to the voltage supply.

Note that the intercept correction factor $\rho_B$ discussed here (which is an approximation to the more general intercept correction factor $\rho_{YX}$ discussed in Ref. 2) is a different physical quantity from the "old" intercept correction factor discussed in papers published prior to 2012. The "old" and "new" factors have different formal definitions, and (in circumstances where both could be applied) would usually have different numerical values. As discussed in Ref. 2, the new intercept correction factor is more cleanly defined and more generally useful than the old one.

For the basic correction factors $\sigma_B$ and $\rho_B$, Ref. 2 provides formulas that can be evaluated by numerical means, as described below. However, for the Schottky-Nordheim (SN) barrier often used in CFE theory, $\sigma_B$ is given by the well-known SN-barrier function $s$, and $\rho_B$ by a simple analytical formula[2] that defines a new SN-barrier function $r_{2012}$. Thus, in the SN-barrier case, it is possible to compare the results of analytical and numerical approaches.

The aims of this paper are twofold. First, it will show, for the SN barrier, that numerical and analytical methods lead to equivalent results. This provides support for the validity of our numerical approach, and constitutes a "proof of concept" for this numerical method. Second, it examines examples of the influence of barrier form on the barrier-form correction factor $\nu_F$, and on $\sigma_B$ and $\rho_B$. We consider a physical effect (correlation-and-exchange) that changes the inner side of the barrier, and another physical effect (field fall-off) that changes the outer side. These explorations are illustrative, and aim to show the general nature of expected effects. Well-established simple models are used here for the electron motive energies. A later task will be to explore these effects for more realistic emitter models.

The structure of the paper is as follows. Section II gives background theory. Section III examines correlation-and-exchange effects for a planar emitter, and Section IV the barrier-form effects that occur with curved emitters. Section V provides a summary and conclusions. The conventions and definitions used in Ref. 2 apply to this paper.

## II. BACKGROUND THEORY

### A. The basic approximation

For convenience, this Section summarizes the high-level enabling theory set out in Ref. 2, and slightly extends discussion of its applicability. Many ways exist of presenting CFE data as FN plots, since any one of several "independent" variables (notably, measured circuit voltage, device voltage, barrier field, scaled barrier field, circuit pseudo-field[5] and device macroscopic field), and any one of several "dependent" variables (notably, current, local current density, and macroscopic current density) can be used. To provide generality, a "universal" theoretical formulation is used here and in Ref. 2, in which $X$ denotes any suitable independent variable, and $Y$ any suitable dependent variable. In this formulation, alternative general forms for the technically complete[6] "universal" FN-type equation are:

$$Y = CX^2 \exp[-G_F] \equiv CX^2 \exp[-v_F S^*/X]. \quad (1)$$

The subscript "F", here and elsewhere, indicates that the parameter refers to a barrier of zero-field height equal to the local work function $\phi$. For this barrier, $G_F$ is the barrier strength (formerly called the JWKB exponent), and $v_F$ ("nu$_F$") is a correction factor related to the barrier's mathematical form. The precise mathematical forms of $S^*$ and $C$ depend on which independent and dependent variables are being used, and (for $C$) on what physical/modelling assumptions are being made; for any particular choice of $X$ the form of $S^*$ is well defined, and $S^*$ is positive in value. In FN coordinates [$\ln\{Y/X^2\}$ vs $X^{-1}$], Eq. (1) becomes a function $L(X^{-1})$: the related FN plot then graphs $L$ against $X^{-1}$. Thus:

$$\ln\{Y/X^2\} = L(X^{-1}) = \ln\{C\} - v_F S^* \cdot X^{-1}. \quad (2)$$

Let $S(X^{-1})$ denote the slope of a FN plot. In field electron emission (FE) measurements, $S(X^{-1})$ is a negative quantity that varies slightly or significantly with $X^{-1}$. The *slope correction factor* $\sigma_{YX}(X^{-1})$ is a positive quantity defined by

$$\sigma_{YX}(X^{-1}) \equiv -S(X^{-1})/S^*. \quad (3)$$

The subscript "*YX*" is a reminder that, in this universal formulation, the value of $\sigma_{YX}(X^{-1})$ depends on the choices made for *X* and *Y*.

There may be series and parallel resistances in a FE measurement circuit, as shown in Fig. 1. If so, the voltage *V* applied to the emitter is not equal to the measured voltage $V_m$, and the current *i* flowing through the emitter is determined by the behavior of the whole circuit, not the emitter alone. In particular, if any leakage current flows parallel to *i*, then *i* will not equal the measured current $i_m$. Leakage currents can usually be eliminated by improved experimental design; thus, theory here assumes leakage current is zero and $i=i_m$. In this case the slope $S(X^{-1})$ of a FN plot derived from the measurements of $V_m$ and $i_m$ is given by

$$S(X^{-1}) = \partial L / \partial (X^{-1}) . \qquad (4)$$

A change is made at this point, from discussion in terms of measured quantities $V_m$ and $i_m$ to discussion in terms of the "universal" variables *X* and *Y*: this is because a widespread practice in the subject area is to use auxiliary equations to "pre-convert" measured variables to other variables (often macroscopic field $F_M$ and macroscopic current density $J_M$) before making FN plots. Consequently it is more convenient to continue the discussion using universal variables.

The above equations yield

$$\sigma_{YX}(X^{-1}) = -[\mathrm{d}\ln\{C\}/\mathrm{d}(X^{-1})]/S^* + \nu_F + X^{-1}\mathrm{d}\nu_F/\mathrm{d}(X^{-1}) + (\nu_F X^{-1}/S^*)\cdot \mathrm{d}(S^*)/\mathrm{d}(X^{-1}) . \qquad (5)$$

Customary practice is to then disregard the first and last terms on the right-hand-side of Eq. (5). This constitutes the *basic approximation*.

When the basic approximation is made, various possible influences on FN plots are thereby ignored. In particular, series resistance in the measuring circuit, field-dependent changes in emitter geometry, and field dependence in work function or emission area, will all contribute to the first or fourth terms in Eq. (5), or both. The basic approximation ignores all such effects.

Explorations of the validity of the basic approximation and of the consequences of invalidity are beyond the scope of the present paper, but will be addressed elsewhere in due course. This approximation is clearly invalid if significant saturation effects occur[5] or if significant geometry changes occur[7,8]. However, these effects apply to some measurements but not to others. Thus, it is helpful to explore the approximation properties, as part of overall exploration of improved FN plot analysis. The basic approximation is in fact better than the orthodox and elementary approximations very often used in the literature, because it can take into account the effects of emitter shape.

In the basic approximation, the basic slope correction factor is denoted by $\sigma_B$ and given by[2]

$$\sigma_B = v_F + X^{-1} dv_F/d(X^{-1}) = v_F - X dv_F/dX = v_F - F_C dv_F/dF_C. \tag{6}$$

The new form ($\rho_{YX}$) of intercept correction factor defined in Ref. 2 allows the tangent to curve $L(X^{-1})$ to be written

$$\ln\{Y/X^2\} = \ln\{\rho_{YX} C\} - \sigma_{YX} S^* \cdot X^{-1}. \tag{7}$$

Hence, if $\rho_{YX}$ can be reliably estimated, then (in the improved tangent method) a value for $C$ can be derived from the measured intercept of an experimental FN plot. In the basic approximation, the basic intercept correction factor $\rho_B$ is given by[2]

$$\ln \rho_B = (\sigma_B - v_F)(b\phi^{3/2}/F) = (\sigma_B - v_F)G_F^{ET}, \tag{8}$$

where $b$ is the Second FN constant[9], $F$ is an appropriate value of local barrier field, and $G_\mathrm{F}^\mathrm{ET}$ $[= b\phi^{3/2}/F]$ is the barrier strength for an elementary triangular (ET) barrier of height $\phi$. For any specific well-behaved barrier form, $\sigma_\mathrm{B}$ and $\rho_\mathrm{B}$ can readily be found by numerical evaluation of barrier-strength integrals, as shown below.

### B.    Numerical evaluation of $\sigma_\mathrm{B}$ and $\rho_\mathrm{B}$

The electron motive energy $M$ is defined by writing the one-dimensional Schrodinger equation for "forwards" motion (i.e., motion towards and normal to the emitter surface) in the form

$$\partial^2\psi/\partial x^2 = (2m/\hbar^2)(E_\mathrm{n} - U)\psi \equiv -\kappa^2 M \psi \qquad (9)$$

where $\psi$ is the electron wave-function, $m$ the electron mass, $\hbar$ Planck's constant divided by $2\pi$, and $\kappa = (2m)^{1/2}/\hbar$. $U$ is the potential energy (PE) in which the electron moves, and $E_\mathrm{n}$ is the component of electron total energy associated with the direction of motion ($x$). The "strength" $G$ of the barrier facing this electron is then defined by the barrier-strength integral

$$G = g_\mathrm{e} \int M^{1/2}\,\mathrm{d}x, \qquad (10)$$

where $g_\mathrm{e}$ $[\equiv 2\kappa]$ is the JWKB constant for an electron[6], and the integral is taken across the barrier (i.e., over the range where $M \geq 0$).

The mathematical "form" of a tunnelling barrier is determined by the specific potential-energy (PE) terms that appear in the mathematical expression for $M$. In the present paper, $M$ includes the zero-field height $H$ of the barrier, an electrostatic term $U^\mathrm{el}$, and a term ($U^\mathrm{ce}$) that represents the correlation-and-exchange PE of the escaping electron. Different expressions are used for $U^{el}$ and $U^{ce}$ in different contexts. This makes $M$ a function of $H$, of the barrier field $F$ (defined as the negative of the electrostatic field at the emitter's electrical surface[10,11]), of distance $x$ measured from the electrical surface, and (where appropriate) of the radius of curvature $r_a$ of the spherical electrical surface used to model an emitter with an approximately spherical apex.

Because the emission is treated as governed by a FN-type equation, interest here is in barriers with $H=\phi$. The related electron motive energy is denoted by $M_F$. From the chosen expression for $M$ (putting $H=\phi$), values are obtained for the barrier strength $G_F$, as a function of barrier field $F$ and other parameters, using Eq. (10). The result for the exact triangular barrier is $G_F^{ET}$, as defined above. The barrier-form correction factor $v_F$ for a barrier of strength $G_F$ is defined by

$$v_F \equiv G_F / G_F^{ET}. \tag{11}$$

It is convenient to define a dimensionless partial differential operator $\mathscr{R}_X[\mathscr{F}(X)]$, which operates on a function $\mathscr{F}(X)$ that is a function of $X$ and other variables, by

$$\mathscr{R}_X[\mathscr{F}(X)] \equiv X^{-1} \partial \mathscr{F} / \partial (X^{-1}). \tag{12}$$

When $X$ is the local barrier field $F$, applying this operator to the barrier strength $G_F$ $[= v_F b \phi^{3/2}/F]$ yields (using Eq. (6)):

$$\mathscr{R}_F[G_F] = (b\phi^{3/2} F^{-1})[v_F + F^{-1} \partial v_F / \partial(F^{-1})] = \sigma_B G_F^{ET}, \tag{13}$$

It follows, using Eq. (8), that

$$\sigma_{\mathrm{B}} = \mathscr{R}_F[G_F]/G_F^{\mathrm{ET}}, \tag{14}$$

$$\ln\rho_{\mathrm{B}} = \mathscr{R}_F[G_F] - G_F, \tag{15}$$

where $\phi$ is to be treated as constant when evaluating $\mathscr{R}_F[G_F]$. The quantities $G_F$ and $\mathscr{R}_F[G_F]$ are easily determined numerically, via the barrier-strength integral (10). It follows that the parameters $v_F$, $\sigma_B$ and $\rho_B$ are easily determined numerically for a barrier of any well-behaved form.

### III. PLANAR EMITTER CORRELATION-AND EXCHANGE EFFECTS

#### A. Introduction

For metal emitters that can be treated as "effectively flat", the outer side of the PE barrier decreases nearly linearly, with slope $-eF$; the inner side is dominated by the correlation-and-exchange (C&E) interaction between the escaping electron and the surface. This C&E interaction is often modeled by Schottky's classical planar image PE $(-e^2/16\pi\varepsilon_0 x)$. In consequence, two barrier models are commonly used in CFE theory: the exact triangular (ET) barrier $M^{\mathrm{ET}}(H,F,x)$, which disregards the image PE, and the Schottky-Nordheim (SN) barrier" $M^{\mathrm{SN}}(H,F,x)$, which includes it. For $H=\phi$, these models are

$$M_F^{\mathrm{ET}}(\phi, F, x) = \phi - eFx, \tag{16}$$

$$M_F^{\mathrm{SN}}(\phi, F, x) = \phi - eFx - e^2/16\pi\varepsilon_0 x. \tag{17}$$

These barriers are shown in Fig. 2, for $\phi$= 4.50 eV.

The ET barrier underestimates the local emission current density (ECD) $J_L$, because both barrier height and barrier width are overestimated. On the other hand, the SN barrier is not physically exact close to the emitter's electrical surface (at $x$=0), because $M_F^{SN}$ diverges to $-\infty$ there. An improved surface barrier was suggested by Cutler and Gibbons[12] (CG). For $H=\phi$, the CG barrier has the form

$$M_F^{CG}(\phi, F, x) = \phi - eFx - e^2/16\pi\varepsilon_0 x + c_{CG} e^2/16\pi\varepsilon_0 x^2 \quad (x>x_s), \tag{18}$$

where $c_{CG}$ is an adjustment parameter (denoted by "$\eta$" in Ref. 12), and $x_s$ is the $x$-value at which the C&E PE for the external electron is cut off, in the case $F$=0.

This model was subsequently used by Cutler and Nagy[13,14] to develop an alternative to the Murphy-Good FN-type equation[15]. Although the physics is slightly better, this alternative has never been seriously used, probably because of its mathematical complexity.

### *B.* *Calculation details*

Calculations have been carried out for material-specific parameters appropriate to tungsten, using the "typical" work-function value $\phi$= 4.5 eV and the CG value (10.3 eV) for the inner PE $\chi$. This value is derived from Table 1 in Ref. 12.

As suggested by CG, a distance $x_1$ is defined by:

$$x_1 = e^2/16\pi\varepsilon_0 \chi \approx 0.035 \text{ nm}, \tag{19}$$

and the adjustment parameter $c_{CG}$ (their "$\eta$") is given by

$$c_{CG} = x_1/4 \approx 0.0087 \text{ nm} . \tag{20}$$

This ensures that (in the absence of any applied field) the CG C&E PE joins smoothly onto the bottom of the conduction band, as shown in Fig. 3. The point of join $x_s$ is given by[12]

$$x_s = x_1/2 \approx 0.017 \text{ nm} . \tag{21}$$

Figure 2 compares the resulting CG barrier with the ET and SN barriers. Clearly, the CG barrier is similar to the SN barrier, except for positions very close to the surface.

## C. Barrier strength and transmission probability

Figure 4 shows how the barrier-form correction factor $v_F$ varies with $1/F$, for $\phi = 4.5$ eV. Although analytical results exist for the ET and SN barriers, numerical evaluation has been applied to all barriers, partly for internal consistency, and partly because it allows consistency checks with analytical treatments of the SN barrier (see below). For the CG barrier, $v_F$ goes to zero (i.e., the barrier vanishes) at a slightly higher field (14.94 V/nm) than for the SN barrier (14.06 V/nm).

Figure 5 shows the corresponding plots of $[-G_F$ vs $1/F]$ (called "Lauritsen plots" in Ref. 16). For the ET barrier the plot is an exact straight line that passes through the origin, and has a slope of about –65 V/nm (for $\phi = 4.5$ eV). As expected[16,17], the plots for the SN and CG barriers lie above and nearly parallel to the plot for the ET barrier. The slight curvature in the SN and CG plots (becoming greater as $1/F$ decreases) can be seen by viewing the plots from an oblique angle.

For the transmission probability $D_F$, a plot of $[\ln D_F$ vs $1/F]$ would be very close to the plot of $[(-G_F)$ vs $1/F]$, except for[16] very small values of $1/F$. However, the behavior of the FN plots of type $[\ln\{J_L/F^2\}$ vs $1/F$] and [$\ln\{i_m/F^2\}$ vs $1/F$], where $J_L$ is the local emission current density, would depend on the size and behavior of the neglected terms in Eq. (5). If these neglected terms were all small, then these plots would lie approximately parallel to the plot of [$(-G_F)$ vs $1/F$], except for very small values of $1/F$.

### D. Basic slope and intercept correction factors

Figure 6 shows basic slope correction factors calculated as described above. Obviously, these are very similar for the SN and CG barriers. At high fields (low $1/F$ values), $\sigma_B$ for the SN barrier ($\sigma_B^{SN}$) cuts off near the expected value of $s(f=1)= 0.833$ (see Table III in Ref. 4); $\sigma_B$ for the CG barrier ($\sigma_B^{CG}$) cuts off at the slightly lower value 0.821.

Figure 7 shows basic intercept correction factors calculated as described above. For the SN barrier, at the cut-off point at low $1/F$ values, the value of $\rho_B^{SN}$ is close to the value 48 predicted analytically for $r_{2012}$ from Eq. (23) in Ref. 2, for $\phi= 4.5$ eV, $f= 1$. Figure 7 shows that the difference in form between the SN and CG barriers causes $\rho_B^{CG}$ to be slightly lower than $\rho_B^{SN}$ (typically $\rho_B^{CG}/\rho_B^{SN}$ is about 0.8). For FN-plot interpretation, this is the most significant difference between the SN and CG barriers. Other factors being equal, the emission-area value extracted using the CG barrier model would be about 20% higher than the value extracted using the SN barrier model.

### E. Comparison of analytical and numerical treatments

Table 1 shows values of various dimensionless parameters used in the theory above, for the illustrative input values $\phi= 4.5$ eV, $F= 5$ V/nm. For the SN barrier, values obtained independently from analytical and numerical approaches are listed; as shown, these values agree to four significant figures (except for $\rho_B$), demonstrating that the two approaches are equivalent and that these particular numerical calculations are sufficiently accurate for practical purposes. (The accuracy could be improved at the cost of longer computation times.)

Table 1. Illustrative values of various dimensionless parameters[a]. For the SN barrier, results from analytical and numerical calculations may be compared.

| Parameter | ET barrier (analytical) | SN barrier (analytical) | SN barrier (numerical) | CG barrier (numerical) |
|---|---|---|---|---|
| $G_F$ | 13.04 | 7.58061 | 7.58054 | 7.797 |
| $\mathscr{R}_F[G_F]$ | 13.04 | 12.2250 | 12.2249 | 12.22 |
| $v_F$ | 1.000 | 0.581270 | 0.581265 | 0.5979 |
| $\sigma_B$ | 1.000 | 0.937397 | 0.937391 | 0.9369 |
| $\ln\{\rho_B\}$ | 0 | 4.64441 | 4.64429 | 4.421 |
| $\rho_B$ | 1.00 | 104.002 | 103.989 | 83.2 |

[a]These calculations are based on the values $\phi = 4.5$ eV, $F = 5$ V/nm, $G_F^{ET} \cong 13.04145$.

## F. *Conclusions relating to correlation-and-exchange energies*

Table 1 shows that the changes in the predicted values of correction factors $v_F$, $\sigma_B$ and $\rho_B$ that result from small changes in the C&E PE component $U^{ce}$ are relatively small. The pre-exponential factor $C$ in Eq. (7) contains[18] a component factor $\lambda_E$ associated with atomic-wave-function effects. The uncertainty in $\lambda_E$ is a factor of 10 or more. The difference in $\rho_B$ by a factor of 0.8 is insignificant in comparison.

Thus, in the present state of theoretical development, there seems little practical merit in incorporating *small* improvements in $U^{ce}$ into the theory of FN-plot interpretation. However, Table 1 also illustrates the well-established fact that significant errors (by a factor of order 100) in the prediction of $\rho_B$ and hence $\rho_{YX}$, can be made if the ET barrier is used as the barrier model when the SN barrier is appropriate. For example, this would generate a significant error in the extraction of emission area. Thus, theories of CFE, and related theories of FN-plot analysis, should always include some appropriate representation of correlation-and-exchange effects, when practical interpretation of the FN plot intercept is attempted.

## IV. CURVED EMITTERS

### A. Introduction

Obviously, real field emitters have a rounded tip. There is a significant body of work relating to the prediction of current-voltage characteristics for rounded emitters of various shapes. In general terms, this body of work predicts that emitter curvature will make the FN plots slightly curved. Provided that the plot curvature is not too great, then the FN plot may be analyzed by fitting a straight line and using calculated slope and intercept correction factors for some chosen emitter shape model. As shown below, this analysis ideally requires independent information about the relevant emitter radius of curvature.

Studies also exist that fit non-linear models to curved FN plots; Ref. 19 is an early example. More recently, Edgcombe and de Jonge have investigated[20,21] a method for extracting information from FN-plot curvature. This is based on fitting a parabola to an experimental FN plot, and represents an alternative approach to that described here. Provided that one can be sure about the physical origins of observed FN-plot curvature, then a curvature-based approach is capable in principle of extracting more information than straight-line fitting. However, as discussed in Ref. 16, its implementation may be complicated in practice, and it may be helpful to carry out straight-line fitting first.

### B. Barriers for curved emitters

To apply the general equations developed in Ref. 2 (and summarized above) to curved emitters, an expression is needed for $M_F$ that includes $\phi$ and terms representing $U^{el}$ and $U^{ce}$. It is well known[22] that both $U^{el}$ and $U^{ce}$ depend on the shape assumed for the electrical surface.

Obviously, the simplest curved shape is a sphere with a smooth classical surface. In this case, the electrostatic term can be written

$$U^{el} = -eFr_a x / (r_a + x),\qquad(22)$$

where $e$ is the elementary positive charge and $r_a$ is the sphere radius. This form is chosen so that $U^{el}=0$ at the sphere surface ($x=0$).

As with a planar emitter, it is customary to model the C&E PE by an image PE. For a sphere held a constant electrostatic potential difference with respect to its surroundings, an exact expression exists[23] for the image PE $U^{im}$, namely

$$U^{im} = -(e^2/16\pi\varepsilon_0 x)\{2r_a/(2r_a + x)\}.\qquad(23)$$

As usual, this form is chosen so that $U^{im}$ (and hence $U^{ce}$) tends to zero as $x$ tends to infinity.

Clearly, for spheres of sufficiently large radius (say >20 nm) this expression represents, in the region of space where the barrier exists (typically in the range 0.5 nm < $x$ < 2 nm), a correction of around 5% or less to the classical planar image PE. In view of the conclusion above that small changes in $U^{ce}$ have only limited effects on correction factors, we have continued to use the planar expression in our calculations. This enables us to assess more clearly the difference made by the change in $U^{el}$. Thus, the motive-energy expression used in our calculations has been

$$M_F = \phi - eFr_a x/(r_a + x) - e^2/16\pi\varepsilon_0 x.\qquad(24)$$

For $\phi$= 4.5 eV as before, calculations have been carried out for a range of barrier fields and for the $r_a$-values 200 nm, 50 nm and 20 nm. Figure 8 shows the barrier shape in these cases. It will be seen that, as $x$ increases, the motive energy $M_F$ tends to "flatten out" towards a limiting value, which is readily shown to be $M_F^{lim} = \phi - eFr_a$.

Mathematically, it is possible to choose combinations of relatively small values of $F$ and $r_a$ such that $M_F^{lim} > 0$. In such cases, the mathematics allows no tunneling. In fact, in such cases, the assumption that the PE variation associated with a sphere is a fair reflection of the true electrostatic PE above a rounded emitter is well past the point of breakdown. When one takes the influence of the shank of a real emitter into account, the electron motive energy does not flatten out in this way, but continues to diminish.

The actual requirement for applicability of the sphere model is that the associated electrostatic PE variation be adequately valid in the region of the barrier - say within 2 nm to 3 nm of the emitter surface. By restricting our choice of radii to 20 nm and above, and our choice of fields to 1 V/nm and above, when the value $\phi$=4.5 eV is used, we remain in the region where the sphere approximation is adequately valid.

Obviously, it is important to have correction-factor values for emitters of much smaller apex radius. To obtain reliable results, the equations developed earlier must be applied to emitter models that take the electrostatic influence of the shank into account, and it may be necessary to use a better model for correlation-and-exchange PE. These things are straightforward (provided that the apex radius is not so small that atomic-level effects come into play). The necessary calculations are considered beyond the scope of the present introductory paper, but the issue is discussed in Section V below.

A second reason for restricting radii values to 20 nm and above in this paper is that this is close to the emitter-size value below which it seems likely to be necessary to discuss the potential influence of quantum-confinement effects[24,25], – which, again, is considered beyond the scope of the present paper.

### C.   *Barrier form correction factor*

Figure 9 shows how $v_F$ varies with $1/F$, for the barrier defined by Eq. 24 and the selected values of $r_a$. For a given value of $1/F$ (where $F$ is the field value at the electrical surface), $v_F$-values for curved emitters are greater than those for a planar emitter, and hence transmission-probability values are less. This is because, for curved emitters, the barrier is slightly wider than the corresponding barrier for a planar emitter (see Fig. 8).

Note that a similar conclusion does not apply to comparisons where the device voltage $V$ is held constant, because the voltage-to-barrier-field conversion factor $\beta_V$ is a function of emitter radius, with $\beta_V=1/r_a$ for a sphere.

### D. Basic slope and intercept correction factors

Figure 10 shows calculated values of the basic slope correction factor $\sigma_B$, for the barrier described by Eq. (30). It is easily seen that, as radius $r_a$ decreases, the rate of variation of $\sigma_B$ with $1/F$ increases. This implies that, as emitter apex radius decreases, the degree of curvature of a FN plot is expected to increase. It is also of interest that, for each radius value, there is a range of $1/F$ values where the plots shown are approximately linear. This suggests that, if the barrier form is the principal cause of FN-plot curvature, then it might be feasible (with careful experiments) to estimate emitter radius from FN-plot curvature (notwithstanding the hesitations expressed in Ref. 16). Figure 11 is a plot of $-G_F$ vs $1/F$, for our chosen $r_a$ values. If the barrier form is the dominant cause of curvature in FN plots, then the curvature exhibited in Fig. 11 will also appear in related FN plots. Certainly for the smaller emitters, the deviation from strictly linearity looks large enough to assess quantitatively.

Figure 12 shows calculated values of the basic intercept correction factor $\rho_B$, for the barrier described by Eq. (30). It is seen that values of $\rho_B$ are much higher for small-radius emitters. The implication is that, even if the barrier form is the principal cause of FN-plot curvature, extraction of reliable values for the parameter $C$ in Eq. (7) (and hence for physical parameters such notional emission area) may require reliable knowledge of the emitter apex radius.

Some years ago, Wang et al.[26] carried out careful experiments on carbon nanotubes (CNTs), including a comparison of emission areas as derived from transmission electron micrographs (TEM) and from FN plots (using planar-emitter theory with a SN barrier). They found that emission areas derived from TEM measurements were significantly less than those derived from FN plots. Although there may be other explanations, this finding is qualitatively compatible with the finding here that $\rho_B$ is much greater for a sharply curved emitter than it is for a planar emitter (thus the extracted emission area would be smaller if curved-emitter theory were used).

As with the planar emitters considered in orthodox data-analysis methods[2,27], accurate interpretation of the slope and intercept of a straight line fitted to a FN plot for a curved emitter requires estimation of the "fitting" barrier field $F_t$ at which the fitted chord is parallel to the tangent to the theoretical FN plot. The numerical results presented in Figs. 10 and 12 show that choice of $F_t$ is likely to be more sensitive for a curved emitter than it is for planar emitter, and that appropriate procedures may need to be developed.

## V. SUMMARY AND CONCLUSIONS

The form of a tunnelling barrier of zero-field height $\phi$ is determined by the related expression for electron motive energy $M_F$. This paper illustrates how barrier form influences the emission parameters associated with FN plots, in particular the basic slope and intercept correction factors $\sigma_B$ and $\rho_B$. There have been many previous theoretical calculations of current-voltage characteristics for curved emitters, but relatively few attempts to consider how to extract information from the related FN plots. Reference 2 set out the general principles of an improved approach. This paper has illustrated this improved approach by applying it to some simple well-defined model barriers. The intention was that the paper should be exploratory, and be able help establish what needed doing next.

For various choices of $M_F$, relevant parameters have been calculated numerically, using Eqs (11), (14) and (15). The good agreement found between analytical and numerical results for the SN barrier, as recorded in Table 1, supports the validity of this numerical approach.

Our main conclusions are as follows. First, if good results are to be achieved, then (as is already well established) the correlation-and-exchange component ($U^{ce}$) of electron potential energy must be represented in $M_F$. However, for metal emitters of moderate to large apex radius, it seems sufficient to represent $U^{ce}$ by Schottky's classical planar image PE. For these emitters there seems limited practical merit in using more detailed expressions for $U^{ce}$ (although this may become necessary for emitters with apex radius smaller than those investigated here).

Second, with a spherical emitter model, it has been found that $\sigma_B$ and $\rho_B$ increase significantly as the emitter radius $r_a$ is decreased. This effect is primarily associated with increase in the rate at which electrostatic field falls off with distance from the emitter surface, as $r_a$ decreases. For a given value of barrier field $F$ (i.e., the negative of the electrostatic field value in the electrical surface), this makes the barrier become thicker as $r_a$ decreases.

It is also found that, at low $F$-values, $\sigma_B$ varies nearly linearly with $1/F$, with the rate of increase of $\sigma_B$ dependent on the value of $r_a$. This offers hope that, with emitters of moderate to small radius, it may be possible to extract the value of $r_a$ by fitting a parabola to the low-field portion of a FN plot, or by determining the plot curvature in some other way. Detailed simulations are now needed, in order to establish the viability of this suggestion.

In order to avoid problems with the physical validity of spherical emitter models, the analysis in this paper has been restricted to $r_a$ values greater than 20 nm. An obvious next step is to apply the numerical approach used here to a more sophisticated model for the electrostatic PE variation above a real field emitter, using a model that remains valid for emitter radii substantially less than 20 nm.

Ideally, one needs to use a model in which the apex radius and effective cone angle of the emitter can be chosen independently. In some historical models, particularly those based on modeling the emitter shape as a hyperbola or a parabola, this is not possible. This suggests that the sphere-on-orthogonal-cone (SOC) model, already used in simulations[28,29] related to field electron emitters, might be a good choice: analytical expressions exist for the electrostatic potential outside it, and the model has three adjustable shape parameters.

For definiteness (and in the interests of minimizing complexity), the calculations here have been set in the context of CFE from a bulk metal emitter. However, barrier aspects of CFE are qualitatively similar for all emitting materials. Hence the discussion here can, up to a point, serve as a model for treating CFE from non-metals and from small emitters, especially when emission from a single band or sub-band is dominant.

For some materials and emitters, however, there will be significant differences in detail. For example, $U^{ce}$ may not be well approximated by the classical image PE for a plane or a sphere, or (when field penetration and band-bending occur) the operative work-function may be a significant function of barrier field and possibly other parameters (such as doping concentrations). Thus, each emission situation needs to be thought about separately.

**Acknowledgements**

Andreas Fischer thanks the Alexander von Humboldt foundation for a Feodor Lynen fellowship and Mu'tah University for the hospitality.

**Figure captions**

**Fig. 1.** Schematic field electron emission measurement circuit, operating in unfavourable conditions. If high series ("$R_s$") and/or low parallel ("$R_p$") resistances exist in the circuit, then the voltage $V$ applied across the emitter, and the emission current $i$, will not necessarily be equal to the measured voltage $V_m$ and current $i_m$.

**Fig. 2.** The three barrier models used in this work to compare correlation-and-exchange effects.

**Fig. 3.** Comparison of approximations for the correlation-and-exchange PE: comparison of Schottky's classical image PE with the improved approximation introduced by Cutler and Gibbons[12] (CG).

**Fig. 4** Variation of the barrier-form correction factor $v_F$ with the reciprocal ($1/F$) of barrier field $F$, for the barrier models indicated.

**Fig. 5.** Variation of the parameter $-G_F$ with $1/F$, for the barrier models indicated.

**Fig. 6.** Variation of the basic slope correction factor $\sigma_B$ with $1/F$, for the barrier models indicated.

**Fig. 7.** Variation of the basic intercept correction factor $\rho_B$ with $1/F$, for the barrier models indicated.

**Fig. 8.** Barrier models used to explore the effects of emitter curvature on emission parameters relating to Fowler-Nordheim-type equations. For algebraic details of models, see text.

**Fig. 9.** Effect of emitter curvature on how barrier-form correction factor $v_F$ depends on $1/F$.

**Fig. 10.** Effect of emitter curvature on how the basic slope correction factor $\sigma_B$ depends on $1/F$.

**Fig. 11.** Effect of emitter curvature on how the parameter $-G_F$ depends on $1/F$.

**Fig. 12** Effect of emitter curvature on how the basic intercept correction factor $\rho_B$ depends on $1/F$.

**Figures (For review only**

**Figure 1.** Schematic field electron emission measurement circuit, operating in unfavourable conditions. If high series ("$R_s$") and/or low parallel ("$R_p$") resistances exist in the circuit, then the voltage $V$ applied across the emitter, and the emission current $i$, will not necessarily be equal to the measured voltage $V_m$ and current $i_m$.

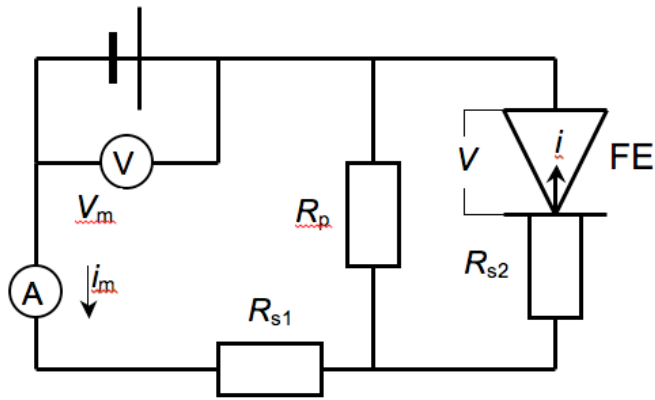

**Figure 2.** The three barrier models used in this work to compare correlation-and-exchange effects.

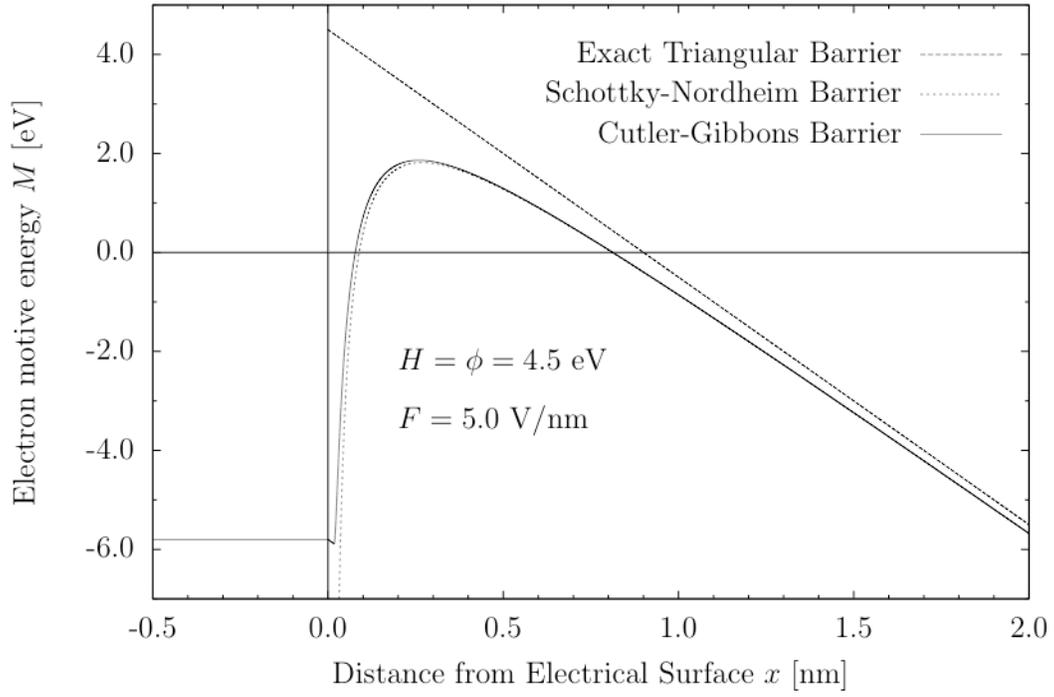

**Figure 3.** Comparison of approximations for the correlation-and-exchange PE: comparison of the Schottky's classical image PE with the improved approximation introduced by Cutler and Gibbons[12] (CG).

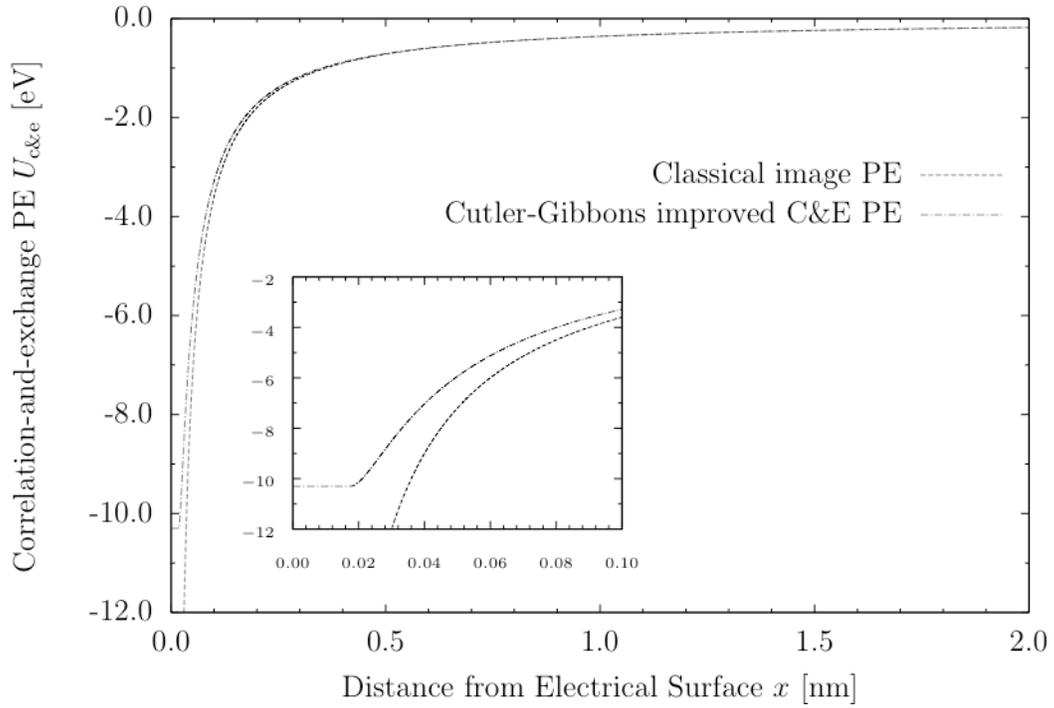

**Figure 4.** Variation of the barrier-form correction factor $\nu_F$ with the reciprocal ($1/F$) of barrier field $F$, for the barrier models indicated.

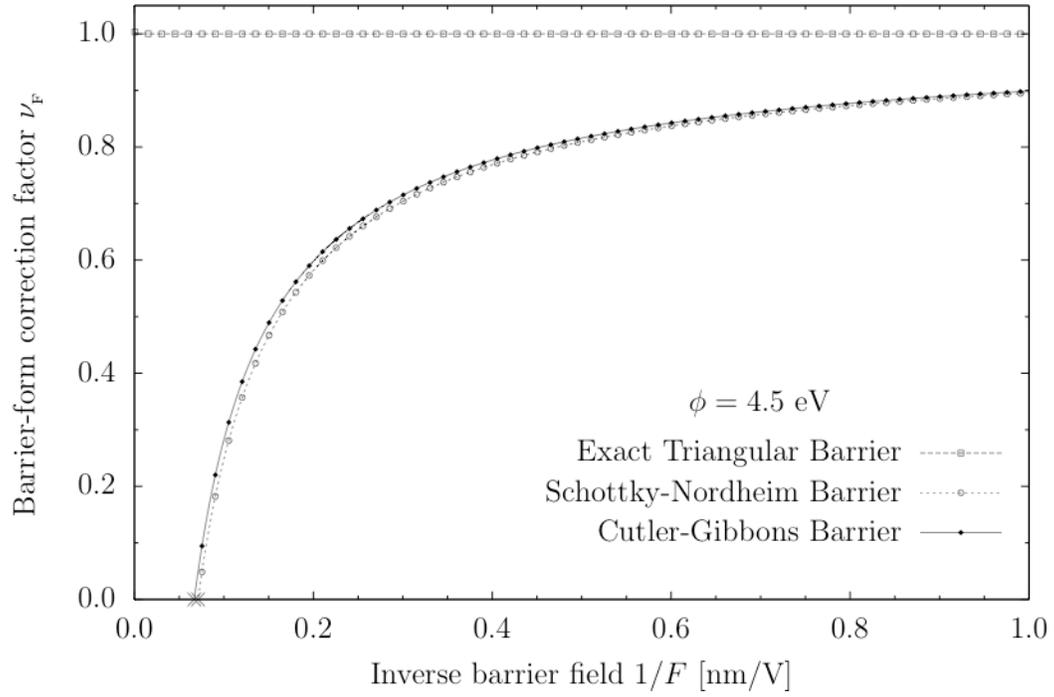

**Figure 5.** Variation of the parameter $-G_F$ with $1/F$, for the barrier models indicated.

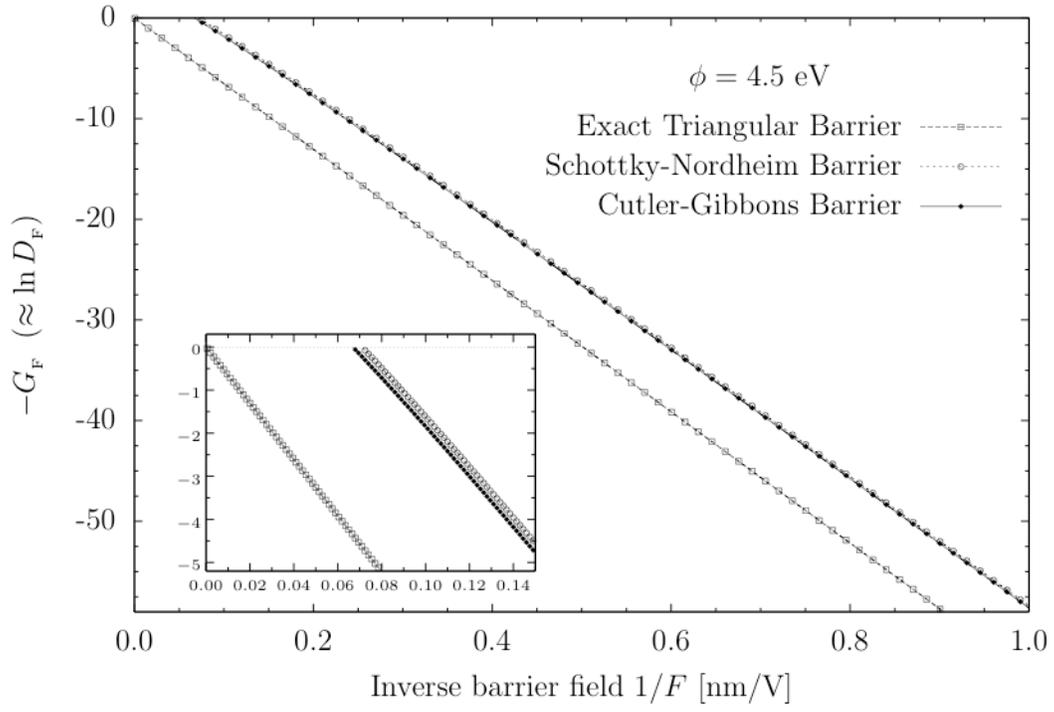

**Figure 6.** Variation of the basic slope correction factor $\sigma_B$ with $1/F$, for the barrier models indicated.

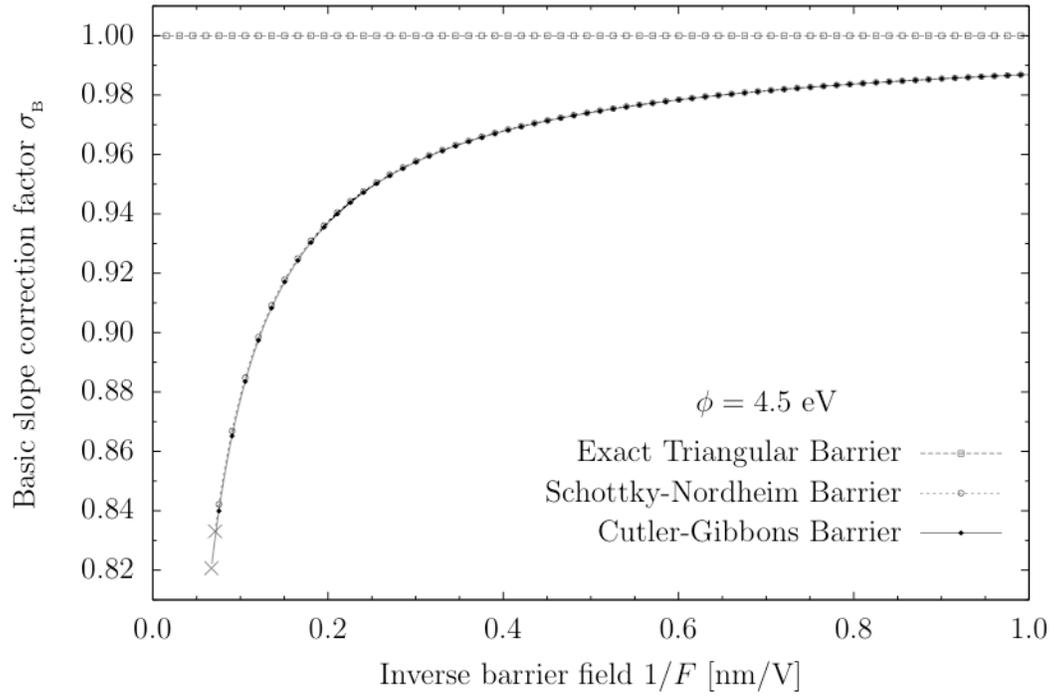

**Figure 7.** Variation of the basic intercept correction factor $\rho_B$ with $1/F$, for the barrier models indicated.

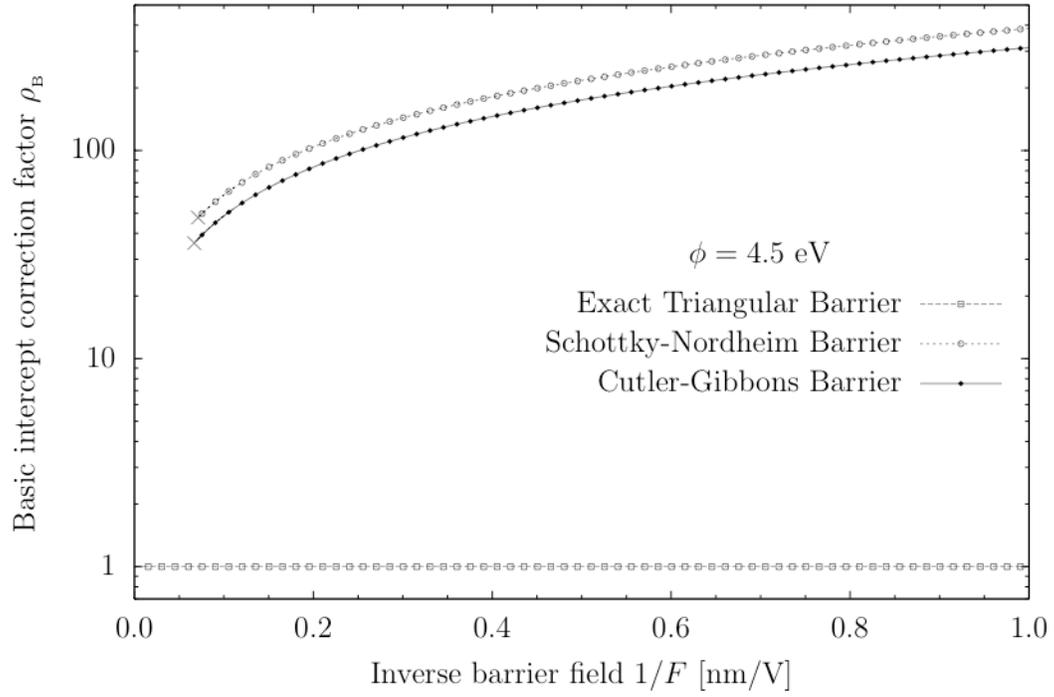

**Figure 8.** Barrier models used to explore the effects of emitter curvature on emission parameters relating to Fowler-Nordheim-type equations. For algebraic details of models, see text.

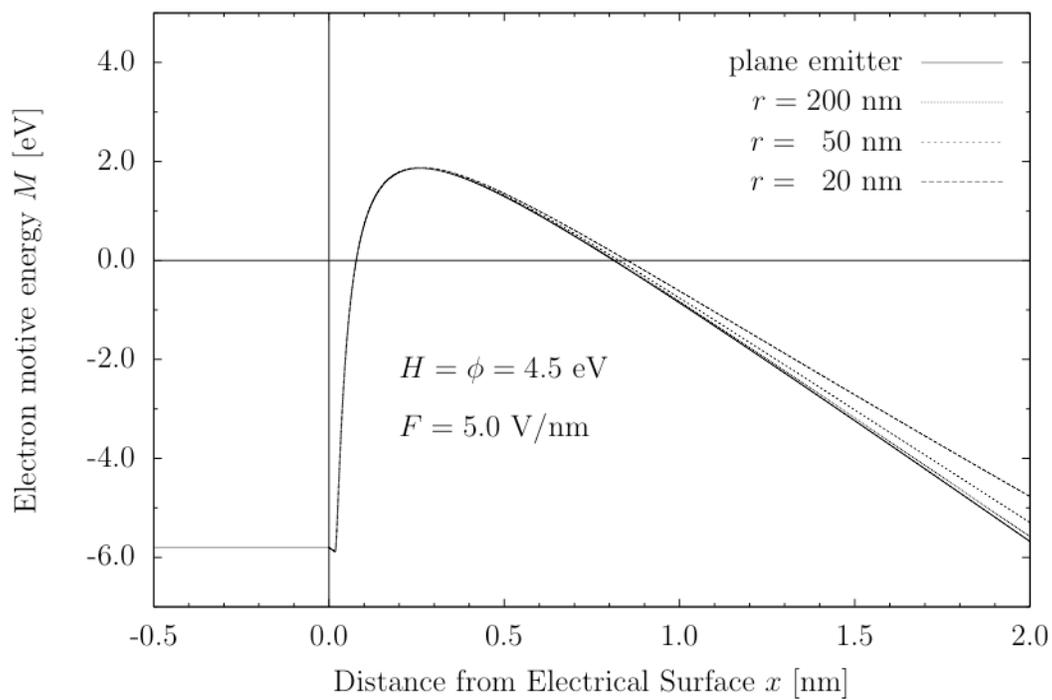

**Figure 9.** Effect of emitter curvature on how barrier-form correction factor $\nu_F$ depends on $1/F$.

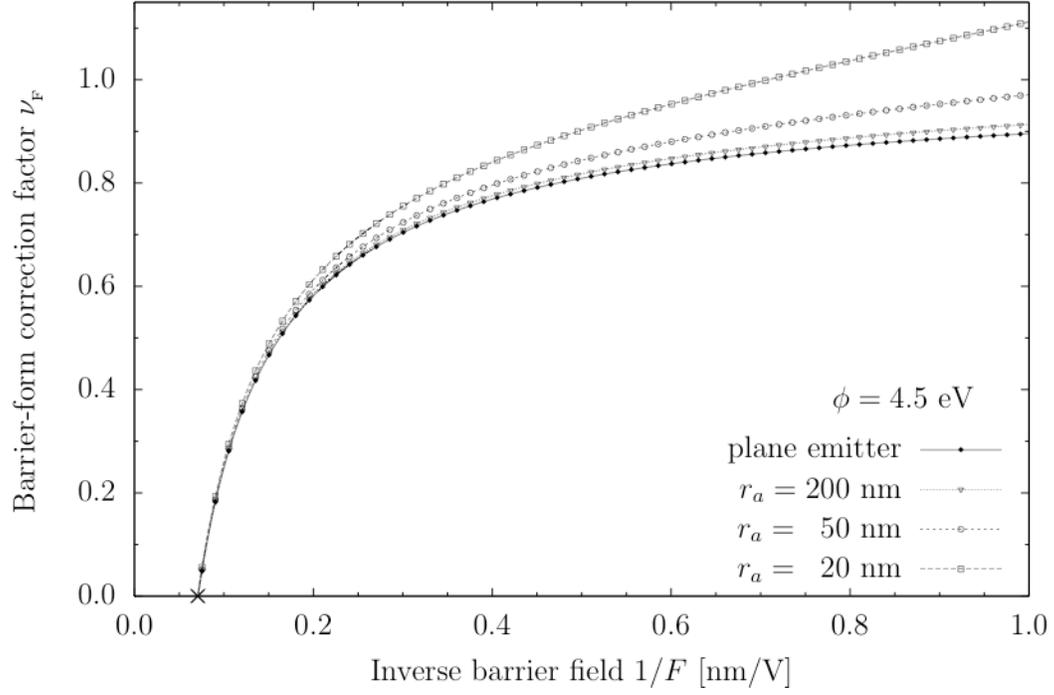

**Figure 10.** Effect of emitter curvature on how the basic slope correction factor $\sigma_B$ depends on $1/F$.

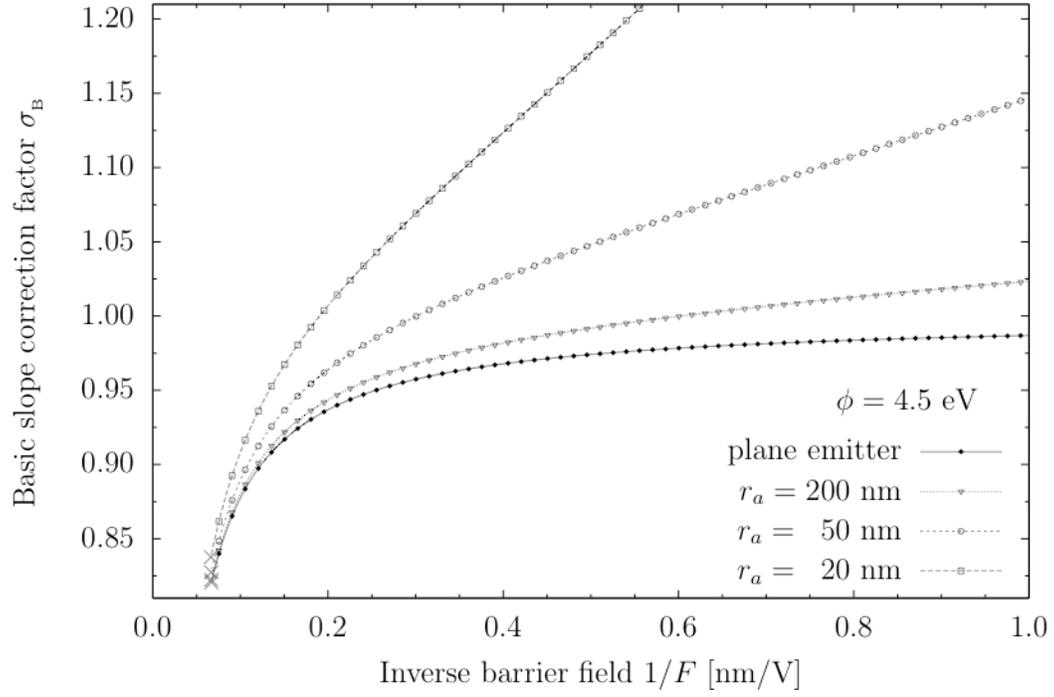

**Figure 11.** Effect of emitter curvature on how the parameter $-G_F$ depends on $1/F$.

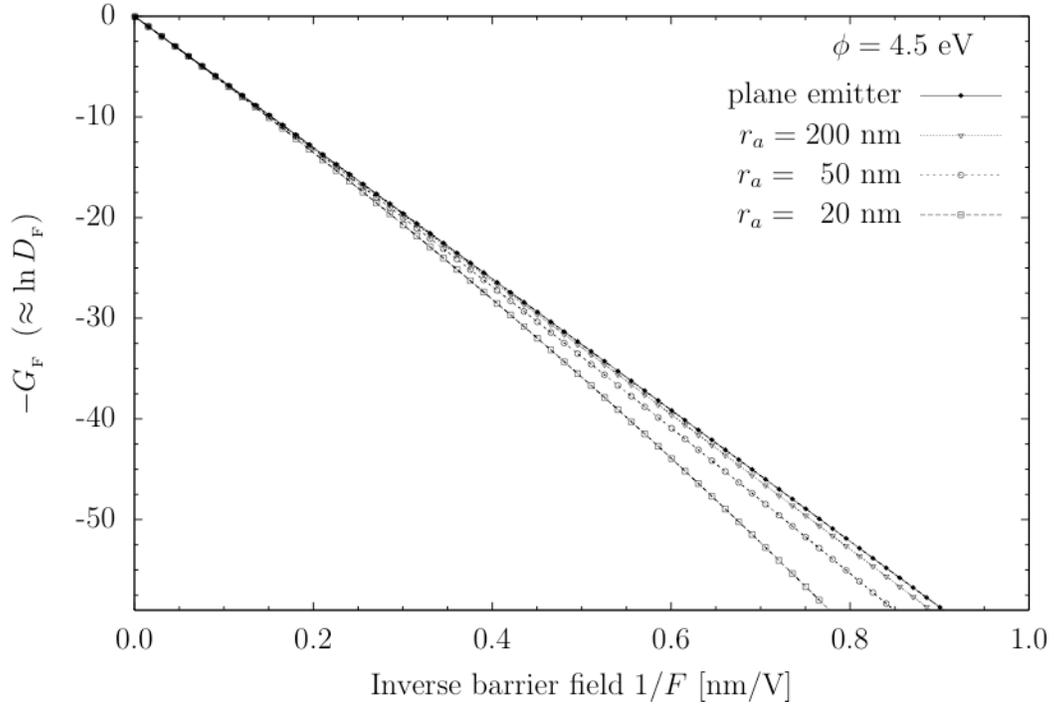

**Figure 12.** Effect of emitter curvature on how the basic intercept correction factor $\rho_B$ depends on $1/F$.

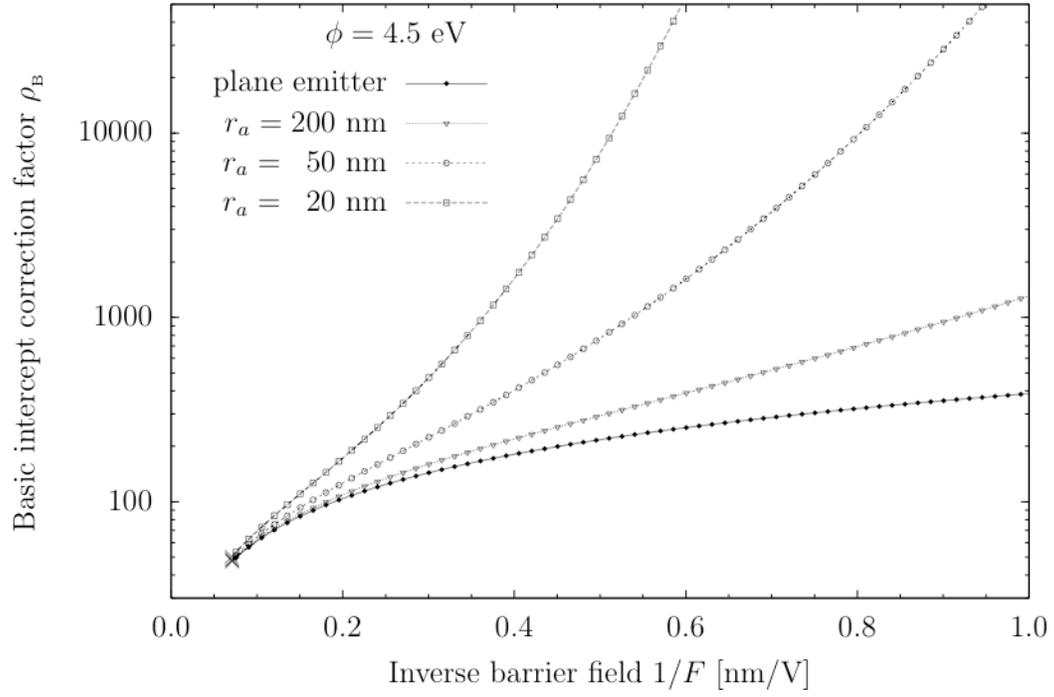